\documentclass[letterpaper]{article}

\usepackage[affil-it]{authblk}

\usepackage[super]{achemso}

\usepackage[T1]{fontenc}
\usepackage[utf8]{inputenc}
\usepackage[margin=1in]{geometry}
\usepackage{setspace}
\usepackage{amsmath,amssymb}
\usepackage{graphicx}
\usepackage{xcolor}
\usepackage{bm}
\usepackage{physics}
\usepackage{siunitx}
\usepackage{mhchem}
\usepackage{float}

\newfloat{scheme}{htbp}{los}
\floatname{scheme}{Scheme}

\newfloat{graph}{htbp}{loh}
\floatname{graph}{Graph}

\definecolor{myred}{HTML}{E76254}

\definecolor{mypink}{HTML}{FF73D0}

\definecolor{mypurple}{HTML}{FF00FF}

\begin{document}

\title{A Photonic Tautochrone: Spatiotemporal Beam Focusing
for Enhanced Nonlinear Response}

\author[a,b,c]{W. Verstraelen\textsuperscript{\S ,}}
\author[a]{S. Zanotti\textsuperscript{\S ,}}
\author[a,b]{N. W. E. Seet}
\author[a]{J. Zhao}
\author[a,d]{D. Sanvitto}
\author[a,b]{J. Zu\~niga-Perez}
\author[a]{K. Dini\textsuperscript{*}}
\author[a,e]{Y. G. Rubo\textsuperscript{*}}
\author[a,b,c]{T. C. H. Liew\textsuperscript{*}}

\affil[a]{
Division of Physics and Applied Physics,
School of Physical and Mathematical Sciences,
Nanyang Technological University,
Singapore 637371, Singapore
}

\affil[b]{
Majulab International Joint Research Unit UMI 3654,
CNRS, Université Côte d'Azur, Sorbonne Université,
National University of Singapore,
Nanyang Technological University, Singapore
}

\affil[c]{
Centre for Quantum Technologies,
Nanyang Technological University,
Singapore 117543, Singapore
}

\affil[d]{
CNR NANOTEC Institute of Nanotechnology,
Via Monteroni, Lecce 73100, Italy
}

\affil[e]{
Instituto de Energías Renovables,
Universidad Nacional Autónoma de México,
Temixco, Morelos 62580, Mexico
}

\date{
\textsuperscript{*}Email:
timothyliew@ntu.edu.sg,
kevin.dini74@gmail.com,
ygr@ier.unam.mx
\\[0.4em]
\textsuperscript{\S}W.V. and S.Z. contributed equally to this work.
}

\maketitle

\begin{abstract}
{We propose to implement an optical analogue of the tautochrone property, which enables isochronous oscillations in mechanical systems, to allow the focusing both in space and in time of ultrashort pulses inside photonic systems. This allows to enhance nonlinear effects, resulting in orders of magnitude increase of nonlinearity-induced phase shifts, while employing low irradiances. Building upon the optical-mechanical analogy, we show that our photonic tautochrone operates as an optical limiter for temporal light pulses, implements temporal bistability and even multistability with large numbers of states, going beyond current bistable optical systems. 
Finally, we extend this concept to the quantum domain and predict a tautochrone quantum blockade regime exhibiting stronger antibunching.}
\end{abstract}

\section*{Keywords}
Nonlinear Optics, Nanophotonics, Quantum Optics, Polaritons, Photon Blockade, Optical Multistability

\maketitle

\section*{Introduction}

In classical mechanics, there is a trajectory along which particles sliding without friction and under constant acceleration all reach the end at the same time, even if they start from different points. This remarkable effect, discovered by Christiaan Huygens, is known as the tautochrone property of the cycloid and has been exploited in the design of accurate pendulum clocks ever since \cite{proctor1878}. In this article we ask whether and how it is possible to implement an analogous tautochrone effect in a photonic system and what effects it would have in the presence of nonlinearity.

Before discussing its possible implementation in photonic systems, we must first address a key difference between mechanics and optics. In mechanics, particles follow a hard slope under a uniform gravitational field, whereas in optics, photons typically travel in straight lines. However, what truly matters for the tautochrone property is not the shape of the path but rather the presence of an effective potential that varies in space. In optical systems, such potential can be created by tailoring the refractive index or by designing the geometry of the medium through which light propagates. From this perspective, photons can experience conditions analogous to those in mechanics, allowing thereby a tautochrone effect to emerge even along straight trajectories.

In order to establish a mathematical equivalence between mechanical and optical systems, we start by describing light propagation in nonlinear media through the nonlinear Schr\"odinger equation \cite{Aranson2002}, also known as the Lugiato-Lefever equation \cite{Lugiato1987}. For photons displaying a parabolic dispersion and undergoing a parabolic potential in the direction transverse to propagation it can be written as:
\begin{align}\label{eq:Schrodinger}
i\hbar\frac{\partial{\psi}({\mathbf{x}},{t})}{\partial t}=
\left(\hbar\omega_c-\frac{\hbar^2{\nabla}^2}{2m}+\frac{1}{2}m\omega^2{\mathbf{x}}^2-\frac{i{\gamma}}{2}+\alpha|{\psi}({\mathbf{x}},{t})|^2\right){\psi}({\mathbf{x}},{t}) +{F}({\mathbf{x}},{t})e^{-i\omega_pt},
\end{align}

This equation describes, for example, the behaviour of light in planar Fabry-Perot microcavities, where light is confined longitudinally along the microcavity direction but spreads in the plane $\mathbf{x}=(x,y)=(r\cos\phi,r\sin\phi$). In this general equation dissipation is accounted for with rate $\gamma$, the nonlinear term represents a Kerr-type nonlinearity, and $F$ represents the amplitude of a coherent driving field (e.g., a laser source).

Let us now consider the optical system driven with an ultrashort laser pulse that is spatially wide, in which case light is injected at different positions in space but the same moment of time. If the pulse varies only slowly in the transverse coordinate, as is typically the case for most laser pulses, the initial energy is dominated by the effective potential, while the kinetic contribution is negligible. This situation is directly analogous to the classical particle placed on the cycloid curve with zero initial velocity.

\begin{figure}[H]
\centering
\includegraphics[width=0.6\columnwidth]{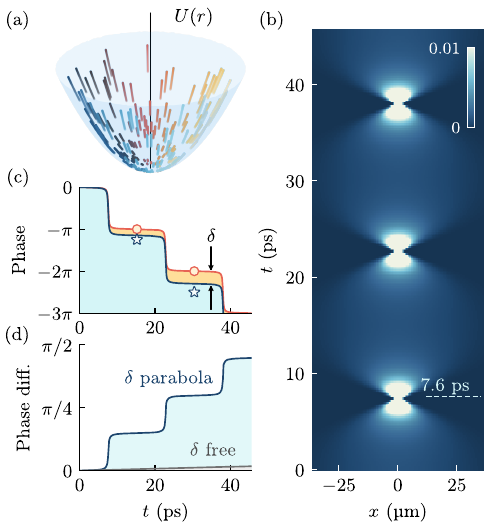}
\caption{{\bf Photonic Tautochrone.} a) A pulsed excitation initializes particles at different positions of a parabolic potential. The particles are then accelerated toward the center corresponding to a focusing of light. b) A space-time plot of intensity at $y=0$ shows that the focusing occurs at $t=T/4$ and integer delays of $T/2$ thereafter. The colour scale has been saturated at $1/100$ the peak value. c) The phase of light in the center undergoes a jump of $\pi$ after each focusing event (red); the shift is enhanced in the nonlinear regime (blue).  Circles and stars mark the times and 
phases of the excitation pulses for the optical limiter and 
bistable regimes, respectively. d) Nonlinearity-induced phase shift for free particles (grey) and in the presence of a parabolic potential (blue). The initial field $\psi(x)$ is normalized to unity. The tautochrone simulations are performed using
$\sigma_0=5l$,
$m=4.22\times10^{-5}m_0$, $m_0$ being the electron mass,
$V(x_{\max})=\SI{3.2}{\milli\electronvolt}$ at
$x_{\max}=\SI{25}{\micro\meter}$,
$\gamma=\SI{0}{\electronvolt}$,
and, for the nonlinear calculations,
$\alpha=\SI{6}{\micro\electronvolt\micro\meter\squared}$.
These parameters correspond to a harmonic-oscillator length
$l=\SI{3.64}{\micro\meter}$ and a level spacing
$\hbar\omega=\SI{0.14}{\milli\electronvolt}$. }.
\label{fig:TautochroneScheme}
\end{figure}

The full analogy becomes complete when realizing that the oscillation frequency of photons in the parabolic potential is independent of their energy, and thus every photon is now accelerated toward the center of the parabolic potential and will reach $\mathbf{x}=0$ at the same time (see Fig.~\ref{fig:TautochroneScheme}a). Thus, in the linear regime and without dissipation, an initial Gaussian wavepacket evolves as $|\psi(t,\mathbf{x})|^2\propto\exp\{-\mathbf{x}^2/(2|\sigma(t)|^2)\}$ and $\sigma(t)$ defines the width with time evolution (see, e.g., Ref.~\cite{Tannor2007} and Supplemental Material for full solution):
\begin{equation}
|\sigma(t)|^2=\sigma_0^2\cos^2(\omega t)+\sigma_1^2\sin^2(\omega t),
\end{equation}
where $\sigma_0=\sigma(0)$ is the initial width, $\sigma_1=l^2/(2\sigma_0)$, and $l=\sqrt{\hbar/(m\omega)}$ denotes the characteristic oscillator length. We note that for a large enough initial width $\sigma_0\gg l$, the wavepacket collapses practically to a point with the width $\sigma_1\ll l$. This behaviour is shown in Fig.~\ref{fig:TautochroneScheme}b. While in theory larger and larger initial widths give arbitrarily tighter focusing, we should note that the paraxial approximation will eventually break down in this simple picture. Overall, a strong focusing effect both in space and time is clearly achieved, reproducing in the optical domain the tautochrone behaviour of beads in the classical mechanical tautochrone system.

We propose that the aforementioned spatiotemporal focusing effect can have strong implications in nonlinear photonics, as it can enhance the effect of particles' interactions in nonlinear media. The article is organized as follows: we first show that the focusing of optical beams can allow orders of magnitude enhancement of nonlinear phase shifts (section {\it Interaction-induced phase-shift---}) as well as analogue spectral shifts (section {\it Optical Limiter---}), which become the building block to implement temporal bistability and multistability, in sections {\it Bistability---} and {\it Multistability---}, respectively. The orders-of-magnitude enhancement of nonlinear effects enabled by the tautochrone mechanism naturally raises the question of whether it can persist in the few-particle regime, where a fully quantum description becomes necessary. This question is explored in detail in section {\it Quantum blockade---}, where we show that the tautochrone effect can indeed shift multiparticle states out of resonance, thereby giving rise to an enhanced quantum blockade regime. Finally, before exploring alternative effective potentials and their associated effects, we present explicit designs of photonic structures that realize the required tautochrone potential, paving the way for experimental demonstrations of the phenomena discussed in this work.

\section*{Results}
\subsubsection*{Interaction-induced phase-shift} While the simplest way to detect interactions in a nonlinear photonic system is to look for an energy shift of the photonic modes involved, the most accurate measurements, which have sensitivity down to the few photon level~\cite{Kuriakose2022}, rely on phase shifts. In Fig.~\ref{fig:TautochroneScheme}c, we find that the phase shift measured in the center of a two-dimensional parabolic potential jumps by $\pi$ in the linear regime every time the particles undergo a focusing event. The phase shift is enhanced in the nonlinear regime and we define the nonlinear-induced phase shift as the difference of the phase in the nonlinear and linear regimes in Fig.~\ref{fig:TautochroneScheme}d. Remarkably, the nonlinear-induced phase shift is over one order of magnitude larger in the parabolic trap as compared to the same pulse without the parabolic potential and the same nonlinear interaction strength.

\subsubsection*{ Optical Limiter}

Aside the prospect of enhanced nonlinear-induced phase shifts, we can also ask if related nonlinear intensity effects would also be enhanced. Among the most fundamental of Kerr nonlinear effects is the optical limiter effect, whereby the injected intensity from a resonant continuous wave laser increases sublinearly with the laser power and which is used to protect devices from overload (including lasers themselves). This effect relies on interactions shifting the photonic field out of resonance with the pumping laser. To achieve this in the tautochrone case is not obvious as we are required to inject all light at the same time with short pulses and, thus, we operate with broadband photonic fields and driving fields. As a solution, we propose a periodic driving of the system with a sequence of ultrashort pulses whose period matches twice the focusing time.

\begin{figure}[H]
\centering
\includegraphics[width=0.6\columnwidth]{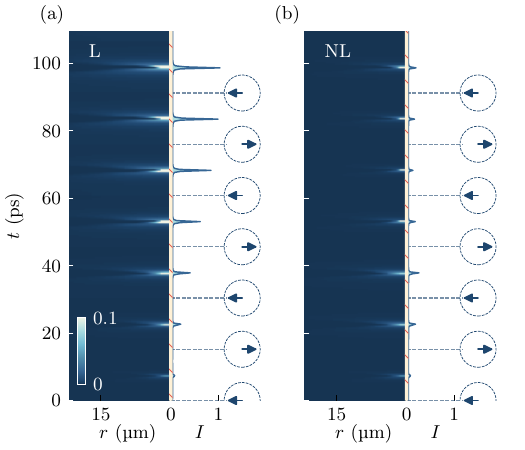}
\caption{{\bf Pulsed Excitation.} 
(a) Intensity $|\psi(r,t)|^{2}$ (left) and its linecut at $r=0$ (right) under periodic excitation without interactions, 
computed from Eq.~\ref{eq:Schrodinger} ($\psi(r,\phi,t)$ is independent of $\phi$, so we do not plot the $\phi$ dependence). The repetition 
period was $T/2$, $\gamma = \hbar/T$, and all other parameters 
were the same as in Fig.~\ref{fig:TautochroneScheme}, with the 
spatially integrated pulse intensity set to unity. Arrows mark 
the pulses arrival times, and their orientation encodes the 
excitation phase; these pulses correspond to the 
circles shown in Fig.~\ref{fig:TautochroneScheme}(c). We have chosen the driving term to have a Gaussian form in space. (b) Same analysis as in (a), but with nonlinear interaction included, which leads to a pronounced quenching of the peaks intensity. Intensities in (a) and (b) have been normalised to the maximum 
peak intensity in (a). For subsequent pulses, the system eventually settles in a state where the intensity is periodic in time.}
\label{fig:PeriodicPulsing}
\end{figure}

In the absence of interactions, we obtain the result shown in Fig.~\ref{fig:PeriodicPulsing}a. Here we have also ensured that subsequent pulses have a phase shift of $\pi$ and are spatially inverted such that each pulse adds constructively to the field and respects $F\big(\mathbf{x},t=(2n+1)T/2\big) =-F(-\mathbf{x},t=nT)$. We account now also for dissipation, considering the regime where the dissipation (lifetime) is comparable to the repetition period. In Fig.~\ref{fig:PeriodicPulsing}b, we find that in the nonlinear regime the intensity is significantly reduced. This is because the nonlinear phase shifts desynchronize the signal from the driving pulses.

\begin{figure}[H]
\centering
\includegraphics[width=0.6\columnwidth]{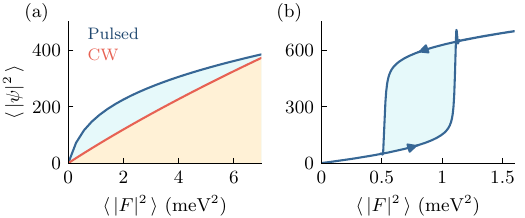}
 \caption{ {\bf Optical Limiter \& Bistability.} a) Power dependence of a photonic tautochrone under periodic pulsed excitation, compared to the same setup under continuous wave (CW) excitation. $\langle|\psi|^2\rangle$ and $\langle|F|^2\rangle$ represent spatially integrated and time-averaged polariton density and driving intensities, respectively. Parameters were the same as in Fig.~\ref{fig:PeriodicPulsing}b. As the overall response to the CW excitation is weaker than the pulsed case, $|F|^2$ is multiplied by a factor of $100$ for CW excitation. b) Same as (a) but with the phase delay between subsequent pulses set to $1.25\pi$. The driving field intensity is slowly ramped up and down, revealing a hysteresis region (shaded).}
\label{fig:OpticalLimiter}
\end{figure}

Figure~\ref{fig:OpticalLimiter}(a) shows that the increase of the photonic field intensity with laser power in the aforementioned pulsed excitation case follows an optical limiter-like curve. This curve appears more nonlinear for the same range of (time-averaged) field intensity when compared to the traditional optical limiter achieved under continuous wave excitation, as represented by the almost linear dependence shown in Figure~\ref{fig:OpticalLimiter}(a)(for a fair comparison between the pulsed and CW regimes, we consider the same parabolic potential and dissipation rate).

\subsubsection*{ Bistability}
Aside the optical limiter effect, a Kerr nonlinear resonator can also demonstrate bistability under continuous wave excitation~\cite{Baas2004} when the driving field is tuned slightly out of resonance. To make use of the tautochrone mechanism, we consider an extension of the phase delay between subsequent pulses exploited in the optical limiter effect as an analogue of the detuning in the continuous wave bistable regime.  Notably, the excitation pulses remain on resonance, in contrast to conventional bistability schemes based on frequency detuning. Figure~\ref{fig:OpticalLimiter}b shows that an extra $0.25\pi$ dephasing of the pumps compared to the optical limiter results in bistability and hysteresis in the system. The physical interpretation is that when there is a low intensity in the system, the focusing and defocusing sequence in the tautochrone is not synchronized with the phases of the repeating driving pulses. However, with some intensity in the system, interactions allow synchronization and maintenance of a high intensity. While traditionally bistability is used to describe a system with two stationary states of the system, we use the term here to describe two spatially and temporally oscillating states that are nevertheless stable and exist under the same conditions. 

\subsubsection*{Multistability} 
\begin{figure}[H]
\centering
\includegraphics[width=0.6\linewidth]{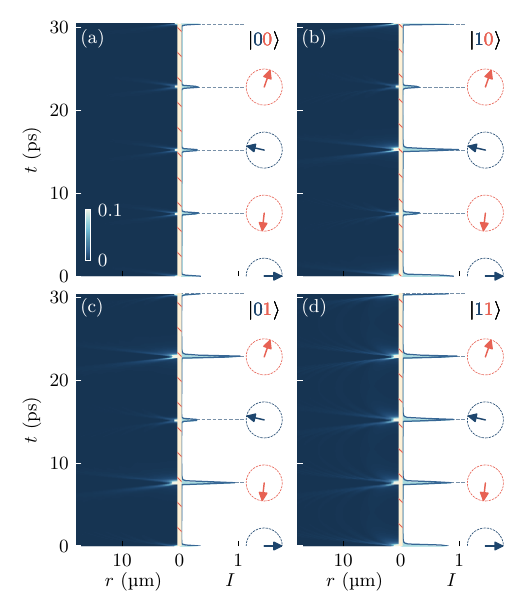}
\caption{{\bf Multistability.}
A photonic tautochrone is driven by two pulse trains with a repetition period of $T/4$, each encoding a different bit. Under these conditions, a hysteresis curve analogous to that in Fig.~\ref{fig:OpticalLimiter}b is obtained, and the pulse intensity is chosen to operate within the multistable regime. Depending on the excitation history, four distinct oscillating  states can be produced. Panels (a–d) correspond to the four possible bit encodings $\ket{00}$, $\ket{01}$, $\ket{10}$, and $\ket{11}$, respectively. 
All remaining parameters are 
the same as in Fig.~\ref{fig:OpticalLimiter}b. The colour scales have been saturated to $1/10$ of the peak 
intensity in (d). Unlike in Fig. 2, here we show the intensity after many pulses such that the initial transient dynamics has been completed.
}
\label{fig:Multistability}
\end{figure}

The aforementioned effect may seem like an over-complicated analogue of already well-established bistability. However, what is quite different to the usual bistability under continuous wave excitation is that the tautochrone mechanism allows generalization to support multiple stable states, with obvious potential applications in communication and information processing. We recall that the sequence of pulses excites an oscillating state that focuses and defocuses periodically. In addition to this state and its associated pulse sequence, we could apply another set of pulses at intermediate times. To minimise cross–interactions between the two pulse sequences, the pulse profile is chosen as a Gaussian multiplied by the square of the radial coordinate, so that the excitation vanishes at the centre. This would effectively enable to double the pulse repetition rate and allow for two states of light oscillating simultaneously in the system, with each state focusing with the arrival of alternate pulses. Figure~\ref{fig:Multistability} shows that four different spatially and temporally oscillating states are now supported under the same conditions.

Interestingly, the multistable regime can be easily generalized to generate a larger number of coexisting stable states. In the Supporting information, instead of doubling the pulse repetition rate, we quadruple it and find that sixteen different stable states are supported under the same conditions.

%
%


\subsubsection*{ Quantum Blockade}

So far, we have discussed a photonic tautochrone and its consequences in the nonlinear regime within the framework of semi-classical theory, including the optical limiter effect. This is appropriate for large numbers of particles where the mean-field approximation offers a good description of coherent states.

Whereas the limiter effect requires higher intensities to be blueshifted out of resonance as $\alpha\abs{\psi}^2$, quantum blockade requires that the nonlinearity already shifts the two-particle quantum state out of resonance, thereby suppressing the probability of having more than one photon inside the cavity at once.
The result is a non-classical antibunched state, characterized by a second order correlation function $g^{(2)}$ dropping below $1$ (for a multiparticle state) and approaching $0$ for a perfect single particle state. For this effect to take place, one typically requires the interaction strength between two particles to exceed the linewidth of the system, giving rise to photon~\cite{Imamoglu1996} or polariton~\cite{Verger2007} blockade effects. While exciton-polaritons are known to achieve nonlinearities substantially stronger than conventional photonics in relatively accessible setups,
only the onset of the polariton blockade, with second order correlations below but still close to $1$, has been attained experimentally~\cite{Delteil2019,MunozMatutano2019} in \ce{GaAs}. There is potential for further enhancement by exploiting specific resonances \cite{Schnuerriger2026}, using materials such as \ce{Cu2O} hosting giant Rydberg excitons \cite{Makhonin2024}, or specifically engineered setups in 2D materials
\cite{Ryou2018}.
Even stronger blockade can be achieved by hybridization of the photon with other matter excitations, atomic states \cite{Gorshkov2011}, or capacitances in superconducting circuits \cite{Carusotto2020}.
Several alternative approaches to antibunching with modest intrinsic nonlinearity, such as unconventional photon blockade and its extensions \cite{Liew2010,Wang2021,Zhou2025,Bamba2011}, have also been proposed. Alternatively, quantum steering \cite{Yuan2025} or reservoir approaches \cite{Ghosh2019,vanregemortel2026} may also generate antibunched states albeit without reaching a steady state.

We note that such single photon states are, beyond fundamental interest, important building blocks for many applications, including quantum technologies such as quantum key distribution and boson sampling, as well as biological research \cite{Couteautechap,Couteauotherap}.

\begin{figure}[H]
\centering
\includegraphics[width=0.6\linewidth]{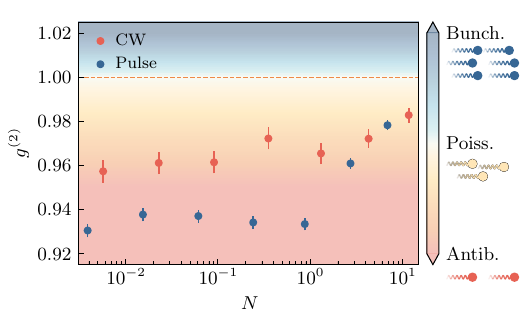}
\caption{{\bf Photon blockade.} For sufficient photon nonlinearity $\alpha$, conventional photon blockade will shift the multi-photon states out of resonance causing antibunching, as is well known under continuous wave (CW) excitation, also in the parabolic trap. The enhancement of interactions through pulses in the tautochrone scheme however, is able to enhance such antibunching. Simulations were done over fixed $\alpha=0.1$~$\omega l^2$, where $l,\omega$ are the harmonic oscillator length and frequency, respectively, while varying $F_0$, which affects also the average particle number $N$. Refer to supplementary for details.
}
\label{fig:G2}
\end{figure}

We now consider a fully quantum theory of our photonic tautochrone, introducing the density matrix $\mathbf{\rho}$ to represent the quantum state of the system, which evolves according to the master equation:
\begin{equation}
i\hbar\frac{\partial{\mathbf{\rho}}}{\partial t}=\left[\mathcal{\hat{H}},\mathbf{\rho}\right]+i\tilde{\gamma}\mathcal{\hat{L}}\left(\mathbf{\rho}\right)
\end{equation}
Here $\mathcal{\hat{H}}$ is the Hamiltonian, which accounts for parabolic kinetic and potential energies together with nonlinear interactions and is defined in the Supporting Information.  The Lindblad superoperator $\mathcal{\hat{L}}$ describes the dissipation of particles at rate $\tilde{\gamma}$. Whereas the conventional quantum blockade is typically calculated for a single quantum mode, we do need to account for spatial dependence explicitly to model the tautochrone behaviour. While this approach results in an intractable size of the density matrix, the master equation can be simulated (to arbitrary accuracy) using a stochastic sampling of the phase space within the Positive P approach \cite{Deuar21} (see the Supporting Information for a full description). 
 The full (equal time) second order correlation function of the trap created by the tautochrone effect is defined as:
\begin{equation} \label{eq:fullgtwo}
g^{(2)}(t)=\frac{\iint\langle \hat{\psi}^\dagger(\mathbf{x},t)\hat{\psi}^\dagger(\mathbf{x}',t)\hat{\psi}(\mathbf{x'},t)\hat{\psi}(\mathbf{x},t)\rangle \dd[2]{\mathbf{x}} \dd[2]{\mathbf{x}'}}{\left(\int\langle \hat{\psi}^\dagger(\mathbf{x},t) \hat{\psi}(\mathbf{x},t)\rangle \dd[2]{\mathbf{x}}\right)^2}
\end{equation}
where $\hat{\psi}(\mathbf{x},t)$ annihilates a particle at position $\mathbf{x}$ and time $t$; the angled brackets denote the expectation value. Figure~\ref{fig:G2} compares the value of $g^{(2)}$ under periodic pulsed excitation to that calculated under continuous wave excitation with the same spatial profile for the same parabolic potential. Although the mechanism of the optical limiter is different under periodic pulsed excitation, we find that it also results in antibunching in the quantum limit. More remarkably, the strength of the antibunching is twice as strong around $N\sim1$. Modest system parameters were assumed, such that the conventional blockade under continuous wave excitation matches experimental reports~\cite{Delteil2019,MunozMatutano2019}. Thus, even using realistic physical parameters, the tautochrone effect displays a clear quantitative advantage and appears as an interesting approach to enhance not only nonlinear effects but also, thanks to them, nonlinear quantum effects. In the next section we describe how to build a photonic system displaying the effect introduced in this work.

\subsubsection*{Designs for Optical Potential Engineering} 
In principle, any photonic system with a spatially varying refractive index or effective refractive index would be suitable for the engineering of a parabolic potential for photons. Exciton-polaritons in microcavities are a good candidate system given that, indeed, parabolic traps have already been realized experimentally some time ago using different techniques to modify the local exciton-potential~\cite{Balili2007,Tosi2012} or the local photonic-potential thanks to a spatially varying cavity thickness~\cite{Besga2015}. In general, exciton-polaritons are renowned for their effective Kerr nonlinearity and there are also other options for potential engineering as described in ~\cite{Schneider2017}. Overall, the current  fabrication state-of-the-art ensures the technological feasibility of microcavities implementing the necessary photonic or polaritonic potentials. 



As a specific example design for implementation of a photonic tautochrone, we choose a planar $\lambda/2$ microcavity (where $\lambda$ is an optical wavelength) with a quantum well in the strong coupling regime. The polaritonic modes of the system experiences a parabolic potential, created using a reservoir of excitons~\cite{Tosi2012,Wei2022}. By spatially modulating a non-resonant laser, the spatial distribution of excitons can be controlled to design a quadratically increasing exciton density. The effective polariton potential is induced by the repulsive Coulomb interaction between the polaritons and the reservoir excitons.

In order to simulate from first principles the dynamics of this tautochrone system in both the linear and non-linear regimes, we use our self-developed Nonlinear Bloch-Maxwell FDTD algorithm \cite{Dini2024,Seet2025}. For simplicity we chose to simulate the behaviour of the 1D tautochrone using a 2D FDTD grid. The reservoir density is assumed to be time-invariant. The exciton dipole moment is assumed to lie entirely in the $xy$ plane of the structure, and the polarization of light is chosen to be transverse electric (TE), with field components $(H_x,E_y,H_z)$. The $y$ direction is assumed to be translationally invariant, $z$ is the growth direction, and $x$ is the direction in which the potential varies. Under these assumptions, the Maxwell-Bloch equations read 

\begin{align}
\nabla \times \mathbf{H} &=  \epsilon \frac{\partial \mathbf{E}}{\partial t}
+  \mathbf{J}_{b} \\[6pt]
\nabla \times \mathbf{E} &= -\mu_0 \frac{\partial \mathbf{H}}{\partial t}, \hspace{0.5cm}\mathbf{J}_{b} = \frac{\partial \mathbf{P}}{\partial t} \\
i\hbar\,\frac{\partial P_y}{\partial t} &=E_{\chi}\,P_y- g E_y+ \alpha (N+N_D) P_y \\[10pt]
\frac{\partial N}{\partial t} &=-2\,\mathrm{Im}\!\left(g E_y P_y^{\ast}\right)\\
\frac{\partial N_D}{\partial t} &=0
\end{align}

where $\mathbf{E}$ and $\mathbf{H}$ are the electric and magnetic fields, respectively;
$\mathbf{P}$ is the material polarization;
$\mathbf{J}_b$ is the bound electric current density;
$\epsilon$ is the spatially varying permittivity;
$\mu_0$ is the vacuum permeability;
$E_{\chi}$ is the exciton resonance energy;
$g$ is the effective light--matter coupling coefficient;
$\alpha$ is the exciton--exciton interaction constant;
$N$ is the spatially varying exciton density; and
$N_D$ is the spatially varying reservoir exciton density responsible for the potential.

\begin{figure}[H]
\centering
\includegraphics[width=0.7\columnwidth]{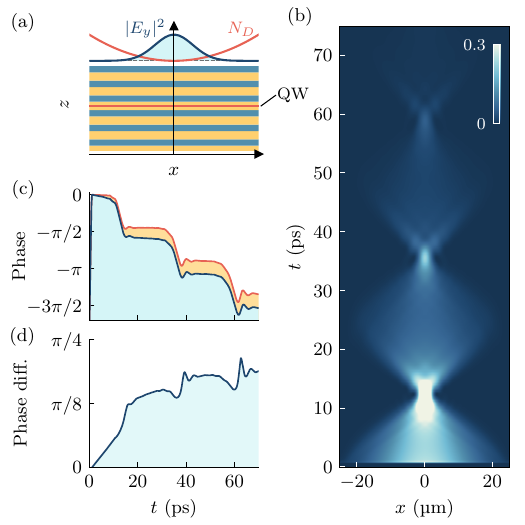}
\caption{{\bf FDTD simulation of a tautochrone.} (a) A quantum well is embedded between two Bragg mirrors, forming a planar microcavity. The system is initialized with a quadratic reservoir population $N_D$, which produces a quadratic potential $V_\chi$. The cavity is excited with a y-polarized Gaussian pump spot. (b) Time evolution of the electric-field intensity 
$|E(x,t)|^{2}$, showing that the focusing signal is lost after 
approximately $\SI{70}{\pico\second}$ due to the finite photon 
dwelling time in the microcavity. The colour scale is saturated 
to $0.3$ of the maximum intensity. (c) Phase of the electric field $E(x=0,t)$ defined in the rotating frame associated with the ground state of the excitonic potential, shown with (red) and without (blue) 
nonlinear interactions. (d) Phase difference between the linear and nonlinear evolutions in the presence of a quadratic potential, showing progressively 
smaller phase jumps at successive focusing events due to the reduced nonlinear interaction as the intensity decays.} 
\label{fig:FDTD_res}
\end{figure}

The microcavity structure, sketched in Fig.~\ref{fig:FDTD_res}a,  consists of a \SI{134.2}{\nano\meter}-thick $\text{Ga}_{0.05}\text{Al}_{0.95}\text{As}$ layer with refractive index $n = 3.05$ , enclosed by distributed Bragg reflectors comprising 20 pairs of $\text{Ga}_{0.05}\text{Al}_{0.95}\text{As}$ and $\text{Ga}_{0.8}\text{Al}_{0.2}\text{As}$ layers with thicknesses of \SI{67.1}{\nano\meter} and \SI{57.6}{\nano\meter}, and refractive indices of 3.05 and 3.55, respectively. A \SI{20}{\nano\meter}-thick GaAs QW ($n = 3.62$) is positioned at the center of the microcavity. The total cavity length is $L = $\SI{50}{\micro\meter}.

We first calculate the dispersion of the system for $N_D = 0$ ~\cite{Dini2024,Seet2025,Shang2025}. The light-matter coupling strength $g$ is chosen to yield a Rabi splitting of \SI{18}{\milli\electronvolt}, defined as the energy separation between the lower and upper polariton branches at zero detuning. The exciton energy is set to $E_{\chi} = \SI{1.50}{\electronvolt}$, while the fundamental cavity mode energy is $E_c = \SI{1.497}{\electronvolt}$,  corresponding to a detuning of \SI{3}{\milli\electronvolt}.

To simulate the pulse dynamics, in analogy with the GPE simulations presented in the previous sections, the optical pump is modelled within the FDTD framework as an incident electromagnetic wave injected from the top boundary and propagating normal to the GaAs microcavity plane.

\begin{align}
    E^{\text{pump}}_y \propto \exp\!\left(-\frac{(t - t_0)^2}{\tau^2}\right)
\exp\!\left(-\,\frac{iE_0}{\hbar}\,(t - t_0)\right) \exp\!\left(-\frac{x^2}{W^2}\right)
\end{align}

where $\tau =$ \SI{200}{\femto\second} is the pulse duration, $W = \SI{20}{\micro\meter}$ is the spot size, $t_0 = 5\,\tau$ is the pulse arrival time and $E_0 = \SI{1.487}{eV}$ is the pulse central energy. To approximate the linear regime, the amplitude of the pump is chosen sufficiently small such that no blueshift is observed, either in the presence or absence of the tautochrone potential. In the nonlinear regime, the pump amplitude is chosen such that, in the absence of the tautochrone potential (i.e., for a depleted reservoir), the resulting blueshift remains smaller than the linewidth. Finally, to realize the tautochrone regime, the reservoir density is then chosen to induce a parabolic exciton blueshift:

\begin{align}
    N_D(x) = \frac{V_\chi^{\text{max}}} {\alpha}  \frac{4x^2}{L^2}.
\end{align}

To remain consistent with experimental reports on GaAlAs microcavities, the maximum exciton blueshift is set to \SI{10}{\milli\electronvolt}~\cite{Tosi2012,Anton2013}.
We observe the same tautochrone effect as in the GPE simulations presented in the previous sections (see Fig.~\ref{fig:FDTD_res} b-d). Unlike those simulations, however, the signal is not reamplified after half a period. Consequently, its intensity gradually decreases, leading to a corresponding reduction of the phase accumulated at each focus.

These simulations provide a first-principles validation, from the photonic perspective, of the Gross–Pitaevskij description and demonstrate that the predicted tautochrone dynamics can be faithfully realized in a photonic microcavity.

\section*{Discussion}
One may wonder whether the aforementioned behavior could also occur with a non-parabolic potential or non-parabolic kinetic energy term. In a classical single-band model, the Hamiltonian for a particle in any one-dimensional potential can be written as a function of position $x$ and momentum $p$. Regardless of the Hamiltonian, it could also be written in terms of the canonical variables $I=(m \omega x^2+p^2/m \omega)/2$ and $\theta=\mathrm{atan}(x/p)$. Here $\theta=\pi/2$ corresponds to initial conditions where particles can have different values of position $x$, but zero momentum $p$; $\theta=\pi$ corresponds to the case where a particle has position $x=0$. Hamilton's equation of motion is $\dot{\theta}=\partial H/\partial I$. The only way for all particles to arrive at $x=0$ at the same time, regardless of their different initial positions (which give different values of $I$), is to choose $H\propto I$, such that $\dot{\theta}$ is independent of $I$. This implies a parabolic potential and parabolic kinetic energy. 
Mathematically, it appears that $H$ could also be supplemented with some function of $\theta$ without breaking the tautochrone behavior; however, such a dependence seems  exotic from a physical point of view.
Note that in a 2D case, similarly strong nonlinear effects could also be promoted by focusing particles on a line (caustic), and a wider class of effective Hamiltonians could lead to such behaviour.

We have focused on the excitation of a photonic tautochrone by a coherent pulsed excitation. One could also ask what happens under a non-resonant gain. Under a simple Landau-Ginzburg type model~\cite{Keeling2008}, which is commonly used in the description of exciton-polariton lasers, we find that an effective optical limiter response can be realized above some threshold. This is quite different to the usual case of continuous wave excitation, where the Landau-Ginzburg model typically results in a linear growth of intensity above threshold. As the effects that we have discussed in this paper take the optical limiter effect as a foundation, we speculate that many of them would also be possible under non-resonant gain as well. 

Finally, although we have considered a spatially uniform dissipation, systems with spatially varying dissipation could offer further advantages. A limitation of the conventional quantum blockade effect is that there is a large uncertainty in the time of generation of a single photon (which can be considered a consequence of the uncertainty principle and monochromatic frequency distribution). In principle, a system could be engineered so that dissipation is strongest at the center of the trap. This would allow a higher probability of emission at the specific times when the polychromatic wavepacket has focused to the trap center.

\section*{Conclusion}
An optical pulse in a medium with an effective parabolic potential becomes focused due to an analogue tautochrone property where all particles reach the center of the potential at the same time. This spatiotemporal focusing results in the enhancement of nonlinear-induced phase shifts by orders of magnitude. Despite the broad frequency range, analogues of the optical limiter effect, bistability, and multistability with large numbers of states can be arranged under periodic pulsed excitation. The periodic focusing of wavepackets results in an increased sensitivity to interactions, which ultimately manifests in a stronger quantum blockade.

\section*{Funding Sources}

SZ, KD, JZP, \& TCHL acknowledge support from the National Research Foundation grant N-GAP (NRF2023-ITC004-001). JZ \& TCHL acknowledge support from the Singapore Ministry of Education (MOE) Academic Research Fund Tier 3 grant ``Quantum Geometric Advantage'' (MOE-MOET32023-0003). WV, DS, \& TCHL acknowledge support from the EU under grant ``Quantum Optical Networks based on Exciton-polaritons'' (Q-ONE). NWES, JZP, DS, \& TCHL acknowledge support from the EU under EIC Pathfinder Open project ``Polariton Neuromorphic Accelerator'' (PolArt, Id:101130304). YGR \& TCHL acknowledge support from DGAPA-UNAM, Grant PAPIIT No. IN108524.

\section*{Supporting Information}

The Supporting Information is available free of charge.

\noindent The SI contains: Dissipation rate constraints; Analytic solution of Eq.~(1) in linear regime; Generalized Multistability; Transformation of Eq. 1 to simplified units; Nonlinear Schr\"odinger Equation with non-resonant gain; 
Quantum tautochrone model; Nonlinear enhancement estimate; From Maxwell to Gross-Pitaevskij equations.

\section*{Notes}
Associated Content: A preprint of this work is available on arXiv.\cite{Verstraelen26}

\section*{Author information}

\subsubsection*{Corresponding Authors}

\noindent\textbf{Kevin Dini} --
\hangindent=2em\hangafter=1
Division of Physics and Applied Physics, School of Physical and
Mathematical Sciences, Nanyang Technological University 637371,
Singapore; Email: kevin.dini74@gmail.com

\vspace{0.5em}

\noindent\textbf{Timothy C. H. Liew} --
\hangindent=2em\hangafter=1
Division of Physics and Applied Physics, School of Physical and
Mathematical Sciences, Nanyang Technological University 637371,
Singapore; Majulab International Joint Research Unit UMI 3654,
CNRS, Université Côte d'Azur, Sorbonne Université, National University
of Singapore, Nanyang Technological University, Singapore;
Centre for Quantum Technologies, Nanyang Technological University,
Singapore; Email: timothyliew@ntu.edu.sg

\vspace{0.5em}

\noindent\textbf{Yuri G. Rubo} --
\hangindent=2em\hangafter=1
Instituto de Energías Renovables, Universidad Nacional Autónoma de
México, Temixco, Morelos, Mexico; Division of Physics and Applied
Physics, School of Physical and Mathematical Sciences, Nanyang
Technological University 637371, Singapore;
Email: ygr@ier.unam.mx

\subsubsection*{Authors}

\noindent\textbf{Wouter Verstraelen} --
\hangindent=2em\hangafter=1
Division of Physics and Applied Physics, School of Physical and
Mathematical Sciences, Nanyang Technological University 637371,
Singapore; Majulab International Joint Research Unit UMI 3654,
CNRS, Université Côte d'Azur, Sorbonne Université, National
University of Singapore, Nanyang Technological University,
Singapore.

\vspace{0.5em}

\noindent\textbf{Simone Zanotti} --
\hangindent=2em\hangafter=1
Division of Physics and Applied Physics, School of Physical and
Mathematical Sciences, Nanyang Technological University 637371,
Singapore.

\vspace{0.5em}

\noindent\textbf{Nathan W. E. Seet} --
\hangindent=2em\hangafter=1
Division of Physics and Applied Physics, School of Physical and
Mathematical Sciences, Nanyang Technological University 637371,
Singapore; Majulab International Joint Research Unit UMI 3654,
CNRS, Université Côte d'Azur, Sorbonne Université, National
University of Singapore, Nanyang Technological University,
Singapore.

\vspace{0.5em}

\noindent\textbf{Jiaxin Zhao} --
\hangindent=2em\hangafter=1
Division of Physics and Applied Physics, School of Physical and
Mathematical Sciences, Nanyang Technological University 637371,
Singapore.

\vspace{0.5em}

\noindent\textbf{Daniele Sanvitto} --
\hangindent=2em\hangafter=1
Division of Physics and Applied Physics, School of Physical and
Mathematical Sciences, Nanyang Technological University 637371,
Singapore; CNR NANOTEC Institute of Nanotechnology, Via Monteroni,
Lecce 73100, Italy.

\vspace{0.5em}

\noindent\textbf{Jesus Zuñiga-Perez} --
\hangindent=2em\hangafter=1
Division of Physics and Applied Physics, School of Physical and
Mathematical Sciences, Nanyang Technological University 637371,
Singapore; Majulab International Joint Research Unit UMI 3654,
CNRS, Université Côte d'Azur, Sorbonne Université, National
University of Singapore, Nanyang Technological University,
Singapore.

\subsubsection*{Author Contributions}

$^*$These authors contributed equally to this work (\textbf{W.V} and \textbf{S.Z}).

\vspace{0.25cm}

 \raggedright


\noindent\textbf{W.V.}: Methodology, Software, Formal analysis, Investigation,
Writing -- Original Draft, Writing -- Review \& Editing.
\noindent\textbf{S.Z.}: Conceptualization, Methodology, Software, Validation,
Formal analysis, Investigation, Writing -- Original Draft,
Writing -- Review \& Editing, Visualization.
\noindent\textbf{N.W.E.S.}: Validation, Software, Writing -- Review \& Editing.
\noindent\textbf{J.Z.}: Methodology, Writing -- Review \& Editing.
\noindent\textbf{D.S.}: Writing -- Review \& Editing, Supervision.
\noindent\textbf{J.Z.P.}: Writing -- Review \& Editing, Supervision.
\noindent\textbf{K.D.}: Conceptualization, Methodology, Software, Validation,
Formal analysis, Investigation, Writing -- Original Draft,
Writing -- Review \& Editing, Supervision, Project administration.
\noindent\textbf{Y.G.R.}: Conceptualization, Methodology, Formal analysis,
Investigation, Writing -- Original Draft,
Writing -- Review \& Editing, Supervision.
\noindent\textbf{T.C.H.L.}: Conceptualization, Methodology, Software,
Formal analysis, Investigation, Writing -- Original Draft,
Writing -- Review \& Editing, Visualization,
Supervision, Funding acquisition.



\vspace{6pt}
\clearpage

\end{document}